\documentclass{ws-procs9x6}
\newcommand{\bea}{\begin{eqnarray}}
\newcommand{\eea}{\end{eqnarray}}
\newcommand{\be}{\begin{eqnarray}}
\newcommand{\ee}{\end{eqnarray}}
\newcommand{\bT}{{\bf b}_\perp}
\newcommand{\bp}{{\bf b}_\perp}
\newcommand{\cal}{\sl}
\begin{document}

\title{GPD's and SSA's}

\author{M. Burkardt}

\address{Department of Physics, New Mexico State University\\
Las Cruces, NM 88011, USA,E-mail: burkardt@nmsu.edu}

\maketitle

\abstracts{
Generalized parton distributions involving transverse polarization
are transversely deformed. The deformation of chirally odd GPDs
is related to a transversity decomposition of the quark angular momentum. 
Potential consequences for T-odd single-spin asymmetries 
(Sivers and Boer-Mulders effects) are discussed.
}

\section{Introduction}	
Hadron form factors provide information about the Fourier transform
of the charge distribution within the hadron. Generalized
parton distributions (GPDs) provide a momentum decomposition
of the form factor w.r.t. the average 
momentum fraction $x = \frac{1}{2}\left(x_i+x_f\right)$ 
of the active quark
\bea
\int dx H_q(x,\xi,t) &=& F^q_1(t)
\quad \quad 
\int dx E_q(x,\xi,t) = F^q_2(t)
\eea
where $F^q_1(t)$ and $F^q_2(t)$
 are the Dirac and
Pauli formfactors, respectively.
$x_i$ and $x_f$ are the momentum fractions of the quark
before and after the momentum transfer.
The momentum direction of the active quark singles
out a direction and it makes a difference
whether the momentum transfer is along this momentum or in a
different direction. GPDs thus not only depend on $x$
and the invariant momentum transfer $t$ but also on the
longitudinal momentum transfer through the variable
$2\xi=x_f-x_i$.

Since
GPDs are the form factor of the
same operator whose forward matrix elements yield the usual
parton distribution functions (PDFs)
\bea
\int \!\frac{dx^-}{2\pi}e^{ix^-\bar{p}^+x}
\left\langle p^\prime \left|\bar{q}\left(-\frac{x^-}{2}\right)
\gamma^+ q\left(\frac{x^-}{2}\right)\right|p\right\rangle
&=& H(x,\xi,\Delta^2)
\bar{u}(p^\prime)\gamma^+ u(p)
\\
+&&\!\!\!\!\!
E(x,\xi,\Delta^2)\bar{u}(p^\prime)
\frac{i\sigma^{+\nu}\Delta_\nu}{2M} u(p).
\nonumber\eea
it is possible to develop a position space
interpretations for GPDs \cite{me:1st}.

\section{Position Space Interpretation for GPDs}
Charge distributions in position space are usually measured
in the center of mass frame, i.e. relative to the
center of mass of the system. For impact parameter dependent PDFs,
the analogous reference point is the $\perp$ 
center of momentum of all partons (quarks and gluons)
$
{\bf R}_\perp = \sum_{i=q,g} x_i {\bf r}_{\perp,i} ,
$
where $x_i$ is the momentum fraction carried by each parton and
${\bf r}_{\perp,i}$ is their $\perp$ position. 
One can
form  eigenstates of ${\bf R}_\perp$
\bea
\left|p^+,{\bf R}_\perp={\bf 0}_\perp,\lambda
\right\rangle
\equiv {\cal N}\int d^2{\bf p}_\perp 
\left|p^+,{\bf p}_\perp,\lambda \right\rangle .
\eea
Impact parameter dependent PDFs are defined using the familiar
light-cone correlation function in such transversely localized
states \cite{me:1st,soper}
\bea\!
q(x,\!{\bf b}_\perp\!) \equiv\! 
\int \!\!\frac{dx^-\!\!\!}{4\pi\!} 
\left\langle p^+\!,{\bf R}_\perp\!\!=\!{\bf 0}_\perp \!\right|
\bar{q}(-\frac{x^-\!\!}{2}\!,{\bf b}_\perp\!)
\gamma^+ q(\frac{x^-\!\!}{2}\!,{\bf b}_\perp\!)
\left|p^+\!,{\bf R}_\perp\!\!=\!{\bf 0}_\perp\!\right\rangle 
e^{ixp^+x^-} 
\eea
and with an additional $\gamma_5$ for the polarized distribution 
$\Delta q(x,{\bf b}_\perp)$.
Impact parameter dependent PDFs are Fourier transforms of GPDs
for $\xi=0$ \cite{me:1st,GPD2}
\bea
\label{master}
q(x, \bT) &=& \int \frac{d^2{\bf \Delta}_\perp}{(2\pi)^2 }
e^{i{\bf \Delta}_\perp\cdot \bT} H(x,0, - {\bf \Delta}_\perp^2)\\
\Delta q(x, \bT) &=& \int \frac{d^2{\bf \Delta}_\perp}{(2\pi)^2 }
e^{i{\bf \Delta}_\perp\cdot \bT} 
\tilde{H}(x,0, - {\bf \Delta}_\perp^2) 
\eea
Due to a Galilean subgroup of $\perp$ 
boosts in the infinite momentum frame 
there are no relativistic corrections to Eq. (\ref{master}).
Furthermore, impact parameter dependent PDFs 
have a probabilistic interpretation very similar (and with the
same limitations) as the usual PDFs\cite{mb:adl}. 
For example, for
$x>0$ (quarks) one finds
$q(x, \bT) \geq \left| \Delta q(x, \bT)\right| \geq 0$.

It is important to utilize theoretical constraints when
parameterizing these functions to supplement
experimental data.
%
%
One such constraint arises directly
from the fact that the reference point for
impact parameter dependent PDFs is the $\perp$ center of momentum.
For $x\rightarrow 1$ the active quark becomes the
center of momentum and therefore $\bT$ can never be large, and 
the $\perp$ width of $q(x,\bT)$ should go to zero for 
$x\rightarrow 1$. For decreasing $x$ the $\perp$ width is expected
to increase gradually. Although the width in the valence region 
should still be relatively compact, its size should increase further once $x$ is small enough for the pion cloud to contribute \cite{weiss}.
Therefore the $t$-dependence of GPDs
should decrease with increasing $x$. This is consistent 
with recent lattice results, which showed that higher moments of GPDs
have less $t$ dependence than lower moments \cite{lattice1}

\section{Transversely Polarized Target}
\begin{figure}
\unitlength1.cm
\begin{picture}(10,3.5)(.5,10.3)
\includegraphics{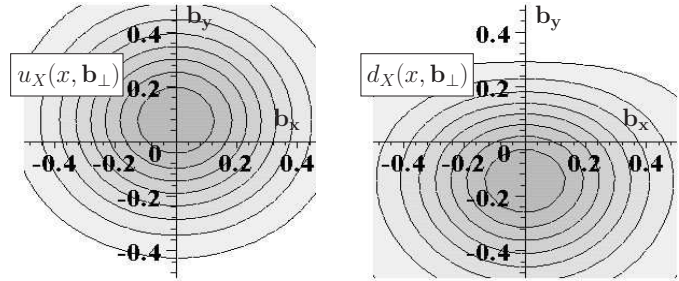}
\end{picture}
\caption{Expected impact parameter dependent PDF for
$u$ and $d$ quarks ($x_{Bj}=0.3$ is fixed) for a nucleon that is 
polarized in the $x$ direction in the model from
Ref. 
For other values of $x$ the distortion looks similar.
}
\label{fig:distort}
\end{figure}

For a $\perp$ 
polarized target, impact parameter dependent PDFs 
are no longer axially symmetric. The deviation
from axial symmetry is described by $E(x,0,t)$. For example, 
the unpolarized quark distribution $q_X(x,\bT)$ for
a target polarized in the $+\hat{x}$ direction reads
\cite{IJMPA}
\be
q_X(x,\!{\bf b_\perp}) = q(x,\!{\bf b_\perp})
-
\frac{1}{2M}\frac{\partial}{\partial b_y}\!
\int \!\!\frac{d^2{\bf \Delta}_\perp }{(2\pi)^2} 
E(x,0,\!-{\bf \Delta}_\perp^2)
e^{-i{\bf b}_\perp\cdot{\bf \Delta}_\perp}.
\label{ijmpa}
\ee
Here $q(x,\!{\bf b_\perp})$ is the impact parameter 
dependent PDF in the unpolarized case (\ref{master}).
This distortion arises since the virtual
photon in DIS couples more strongly to quarks that move
towards it than quarks that move away from it (hence the
$\gamma^+$ in the quark correlation function relevant for DIS).
\cite{IJMPA,ji+}
If  the orbital motion of the
quarks and the spin of the target are correlated
then quarks are more likely
to move towards the virtual photon on one side of the target than 
the other and the distribution of quarks in impact
parameter space appears deformed towards one side. The details
of this deformation for each quark flavor
are described by $E_q(x,0,t)$, which is not known yet.
However, sign and overall scale can be estimated by considering
the mean displacement of flavor $q$
($\perp$ flavor dipole moment) 
\be
\label{dipole}
d^q_y \equiv \int\!\! dx\!\int \!\!d^2\bp
q(x,\bp) b_y
=\frac{1}{2M} 
\int\!\! dx E_q(x,0,0) = \frac{\kappa_{q}^p}{2M} .
\ee
$\kappa_q={\cal O}(1-2)$ 
are the  contributions
from each quark flavor to the anomalous magnetic moment of the
nucleon, i.e.
$F_2(0) = \frac{2}{3} \kappa_u - \frac{1}{3}\kappa_d-\frac{1}{3}\kappa_s...$, yielding $\left|d^q_y\right| ={\cal O}(0.2 fm)$, with opposite 
signs for $u$ and $d$ quarks (Fig. \ref{fig:distort}).

The $\perp$ distortion can also be linked to Ji's relation\cite{ji}
between the $2^{nd}$ moment of the GPDs
$H_q$ and $E_q$ and the quark angular momentum
\be
J_q^i=\varepsilon^{ijk} \int d^3 x M_q^{0jk}.
\label{J}
\ee
Here $M_q^{0jk} = T_q^{0k}x^j-T_q^{0j}x^k$ and
\be
T^{\mu \nu}_q = i\bar{q} \gamma^\mu \stackrel{\leftrightarrow}{D^\nu} q
\label{T}
\ee
(a symmetrization in $\mu$ and $\nu$ is implicit). Since the
angular momentum is obtained by taking the weighted average
of the position, where the weight factor is the momentum density,
one would intuitively expect some connection between the
transverse center of momentum for the quarks and their angular
momentum. Indeed, as has been shown in Ref. \refcite{mb:odd}, 
one can relate the $\perp$ shift of the
center of momentum for a quarks with flavor $q$ to the
angular momentum carried by these quarks. Using Eq. (\ref{ijmpa})
and taking into account an overall $\perp$ shift due to boosting
to the infinite momentum frame one thus recovers the
Ji relation \cite{ji}
\be
\langle J^i_q\rangle = S^i \int dx\, x\left[H_q(x,0,0)+E_q(x,0,0)\right],
\label{JiSR}
\ee
where $S^i$ is the nucleon spin. In combination with measurements
of the fraction of the quark spin contribution to the
nucleon spin 
in polarized DIS, Eq. (\ref{JiSR}) is expected to provide novel
information about the orbital angular momentum carried by the quarks.

The deformation of quark distributions in a $\perp$ polarized
target also provides a very physical source for single-spin
asymmetries (SSA) in semi-inclusive DIS. 
The Sivers function $f_{1T}^{\perp q}$, 
which parameterized the left-right asymmetry 
reads \cite{sivers,trento}
\bea
f_{q/p^\uparrow}(x,{\bf k}_\perp)&=&f_1^q(x,{\bf k}_\perp^2)
-f_{1T}^{\perp q}(x,{\bf k}_\perp^2)\frac{(\hat{\bf P}\times 
{\bf k}_\perp)
\cdot{{\bf S}}}{M} ,
\label{S}
\eea
where $f_{q/p^\uparrow}(x,{\bf k}_\perp)$ represents the
unintegrated parton density for quarks ejected from a $\perp$
polarized target.
The phenomenology of these functions can be found for example 
in Ref.\refcite{ansel} and references therein.
Although one may naively expect that these T-odd functions
vanish, they  survive the Bjorken
limit due to final state interactions 
\cite{BHS,ji2,collins}.

\begin{figure}
\unitlength1.cm
\begin{picture}(10,1.8)(3.,19.2)
\includegraphics{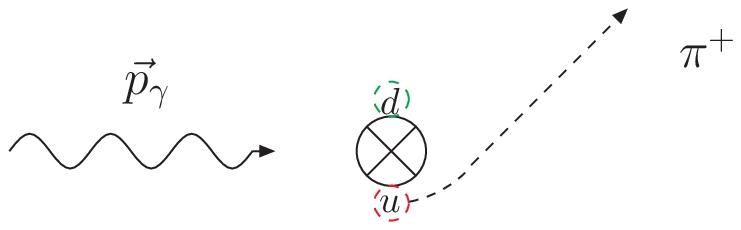}
\end{picture}
\caption{The transverse distortion of the parton cloud for a proton
that is polarized into the plane, in combination with attractive
FSI, gives rise to a Sivers effect for $u$ ($d$) quarks with a
$\perp$ momentum that is on the average up (down).}
\label{fig:deflect}
\end{figure}

For an (on average) attractive final 
state interaction, the position
space deformation into the $+\hat{y}$ direction translates
into a momentum space asymmetry for the ejected quark that prefers
the $-\hat{y}$ direction and {\it vice versa} 
(Fig. \ref{fig:deflect})
Since the sign of the position space distortion is governed by
the sign of the anomalous magnetic moment contribution $\kappa_{q/P}$
from each
quark flavor, this implies that the sign of the SSA is correlated
to the sign of $\kappa_{q/P}$. Following the Trento convention,\cite{trento} this yields a negative Sivers
function $f_{1T}^{\perp u}$  in the proton, while
$f_{1T}^{\perp d}>0$.\cite{mb:siv}
For neutrons the signs are reversed. These predictions 
are consistent with recent HERMES
data\cite{HERMES}. \label{sec:sivers}

\section{Chirally Odd GPDs}
The distribution of transversely polarized quarks in impact 
parameter space is described by the Fourier transform of
chirally odd GPDs \cite{dh}. For an unpolarized
target the distribution of quarks with transversity $s^i$
reads
\be\!\!\!\!\!\!
q_i(x,\!\bT\!)\!= \!-\frac{s^i\varepsilon^{ij}}{2M} 
\frac{\partial}{\partial b_j} \!\!\int\!\! \frac{d^2{\bf \Delta}_\perp}{(2\pi)^2}\!
\!\left[ 2{\tilde H}_T(x,\!0,\!-{\bf \Delta}_\perp^2\!)
\!+ \!E_T(x,\!0,\!-{\bf \Delta}_\perp^2\!)\right]\!\!
e^{-i\bT\!\cdot{\bf \Delta}_\perp}
\label{dh}
\ee
While Eq. (\ref{ijmpa}) describes the $\perp$ deformation
of unpolarized quark distributions in a $\perp$ polarized nucleon,
Eq. (\ref{dh})  demonstrates that
a similar deformation is present in the distribution
of $\perp$ polarized quarks in an unpolarized nucleon ---
except the latter deformation is described by the chirally odd
GPDs $2\tilde{H}_T+E_T$. In Sec. \ref{sec:sivers}, we 
linked the $\perp$ deformation
of the unpolarized quark distributions in a $\perp$ polarized
nucleon to the angular momentum carried by those quarks, yielding 
the Ji relation (\ref{JiSR}) which tells us how the quark angular momentum
is correlated to the nucleon spin. Intuitively, we thus expect
that there is some connection between chirally odd GPDs,
which describe the $\perp$ deformation of $\perp$ polarized
quark distributions in an unpolarized nucleon, and the correlation
between the quark spin and the quark angular momentum in an 
unpolarized nucleon.

In order to investigate the correlation between
polarization and angular momentum of the quarks, we  
decompose $J_q^x$ into transversity
components. 
The projector  on $\perp$ spin (transversity)
eigenstates $P_{\pm \hat{x}} \equiv \frac{1}{2}\left( 1 \pm 
\gamma^x\gamma_5 \right)$ 
commutes with $\gamma^0$, $\gamma^y$, and $\gamma^z$. Hence
all components of the energy momentum tensor
that appear in the definition of $J_q^x$ do not mix between 
transversity
(in the $\hat{x}$ direction) states, defined as
$
q_{\pm\hat{x}} =  \frac{1}{2}\left( 1 \pm 
\gamma^x\gamma_5 \right) q
$.
It is thus possible to decompose
$J_q^x$ into transversity components
\be
J^x_q = J^x_{q,+\hat{x}} + J^x_{q,-\hat{x}} .
\ee
Transversity projections of Eq. (\ref{J}) yield the
transversity components $J^x_{q,\pm\hat{x}}$ 
\be\!\!\!\!\!\!
J^x_{q,\pm\hat{x}} =\!\frac{i}{2}\!\int \!\!d^3x \bar{q}_{\pm\hat{x}}
\!\left[\gamma^0 \!\stackrel{\leftrightarrow}{D^z} +\gamma^z 
\!\stackrel{\leftrightarrow}{D^0}
\right] \!q_{\pm\hat{x}}\, y \, - \! ``y\leftrightarrow z''
= \frac{1}{2}\!\left[ J^x_{q} \pm \delta^x J^x_{q}\right],
\label{Jodd1}
\ee
where the transversity dependent piece reads
\be
\delta^x J^x_{q} = \frac{1}{2}\int d^3x \bar{q}\!\left[
\sigma^{0x} \stackrel{\leftrightarrow}{D^z} + \sigma^{zx}
\stackrel{\leftrightarrow}{D^0}\right]\! q \, y \, - \, ``y\leftrightarrow z''.
\label{Jodd2}
\ee
Taking the matrix elements of $J^x_{q}$ yields the Ji relation 
(\ref{JiSR}).
In order to examine the contribution from the chirally odd term
(\ref{Jodd2}), we consider the form factor of the transversity
density with one derivative \cite{dh,DH2}
\bea
\left\langle p^\prime \right| \bar{q}\sigma^{\lambda \mu}
\gamma_5 i\!\stackrel{\leftrightarrow}{D^\nu} q \left| p
\right\rangle &=& \bar{u} \sigma^{\lambda \mu} \gamma_5 u\,
\bar{p}^\nu A_{T20}(t) + \frac{\varepsilon^{\lambda \mu \alpha
\beta} \Delta_\alpha 
\bar{p}_\beta \bar{p}^\nu}{M^2}\bar{u}u \,\tilde{A}_{T20}(t)
\label{formfactor}\\
+& &
\!\!\!\!\!  \frac{\varepsilon^{\lambda \mu \alpha
\beta} \Delta_\alpha  \bar{p}^\nu}{2M}
\bar{u}\gamma_\beta u \,B_{T20}(t)
+ \frac{\varepsilon^{\lambda \mu \alpha
\beta} \bar{p}_\alpha \Delta^\nu}{M}\bar{u}\gamma_\beta u \,
\tilde{B}_{T21}(t). \nonumber
\eea
Antisymmetrization in $\lambda$ and $\mu$ and
symmetrization in $\mu$ and $\nu$ is implied. 
The 
form factors in Eq. (\ref{formfactor}) are the $2^{nd}$ moments of
chirally odd GPDs
\bea\!\!
A_{T20}(t) &=& \int_{-1}^1 \!\!dx x H_T(x,\xi,t) \label{second}\quad
\quad \quad \quad
\tilde{A}_{T20}(t) = \int_{-1}^1 \!\!dx x \tilde{H}_T(x,\xi,t) 
\\
\!\!B_{T20}(t) &=& \int_{-1}^1 \!\!dx x E_T(x,\xi,t) 
\quad \quad \quad
-2\xi \tilde{B}_{T21}(t) = \int_{-1}^1 \!\!dx x \tilde{E}_T(x,\xi,t) .
\nonumber
\eea
The chirally odd GPDs entering Eq. (\ref{second}) 
are defined as non-forward matrix elements of light-like
correlation functions of the tensor charge
\bea
\!\!\!\!&p^+&\!\!\!\int \!\frac{dz^-}{2\pi} e^{ixp^+z^-}\!
\left\langle p^\prime \right|
\bar{q}\left(-\frac{z}{2}\right) \sigma^{+j}\gamma_5
q\left(\frac{z}{2}\right)\left|p\right\rangle\,
= H_T(x,\xi,t) \bar{u} \sigma^{+j}\gamma_5u +
\\
& &\!\!\!\!\!\!\!\!\!\!\!\tilde{H}_T(x,\xi,t)\varepsilon^{+j\alpha\beta}
\bar{u}\frac{\Delta_\alpha p_\beta}{M^2}u
+\varepsilon^{+j\alpha\beta} \!\!\left[{E}_T(x,\xi,t)
\bar{u}\frac{\Delta_\alpha \gamma_\beta}{2M}u
+\!\tilde{E}_T(x,\xi,t)
\bar{u}\frac{p_\alpha \gamma_\beta}{M}u\right]\nonumber
\eea
Upon taking the expectation value of $\delta^xJ^x$ in
a delocalized wave packet (rest frame), 
the factor $y$ ($z$) projects out terms linear
in $\Delta^i$ in Eq. (\ref{formfactor})
\be
\langle \delta^x J^x\rangle = \frac{1}{2} \int dx \,x 
\left[H_T(x,0,0)+2\tilde{H}_T(x,0,0) + E_T(x,0,0)\right].
\label{SR1}
\ee
yielding a decomposition of the Ji relation into transversity
components
\be
\langle J^x_{q,\pm \hat{x}} \rangle
&=& \frac{S^x}{2}\int dx \,x
\left[H(x,0,0) + E(x,0,0)\right]\label{SR2}
\\
&\pm& \frac{1}{4} \int dx \,x
\left[H_T(x,0,0)+2\tilde{H}_T(x,0,0) + E_T(x,0,0)\right].
\nonumber
\ee
Here $S^x$ is the spin of the nucleon and for an unpolarized
target, only the
second term contributes.
Although the derivation presented above was for one specific
component, it is evident that
rotational invariance implies analogous relations
for $J^y_{q,\pm \hat{y}}$ and $J^z_{q,\pm \hat{z}}$.
Similar relations can also be derived for a spinless target, 
such as a pion.
The scale dependence of
Eq. (\ref{SR1}) is the same as for 
the second moment of the quark transversity
$\int dx\, xH_T(x,0,0)$.

It is instructive to apply our new relations to a point-like
spin-$\frac{1}{2}$ particle, where the ``quark'' spin is always 
equal to the ``nucleon'' spin $H(x,0,0)=H_T(x,0,0)=\delta(x-1)$
and $E=\tilde{H}_T=E_T=0$. In this case 
$\langle\delta^x J^x\rangle=\frac{1}{2}$.
For an unpolarized target there is a $50\%$ probability that 
$S^x=\frac{1}{2}$ and a $50\%$ probability that 
$S^x=-\frac{1}{2}$. When a quark has $s^x=\frac{1}{2}$, which
occurs with $50\%$ probability, the quark also
has $J^x=\frac{1}{2}$, resulting in 
$\langle J^x(s^x=+1/2)\rangle = 0.5 \times \frac{1}{2} = \frac{1}{4}$,
which is consistent with Eqs. (\ref{SR1},\ref{SR2}).
As a second example, when the same point-like ``nucleon'' 
has $S^x= +\frac{1}{2}$, all
of its angular momentum is carried by ``quarks'' with 
$s^x = +\frac{1}{2}$, while none is carried by ``quarks'' with 
$s^x = -\frac{1}{2}$. This is again consistent with Eq. (\ref{SR2}).
A constituent quark model estimate for Eq. (\ref{SR1}) can be found in
Ref. \refcite{pasquini}.

In the general case, when the second moments of the involved GPDs are
nontrivial, it is expected that Eqs. (\ref{SR1}) and (\ref{SR2}) 
will provide novel insights about the spin structure and spin-orbit
correlation for quarks in the nucleon. While experimental results
for $H_T(x,0,0)$ are expected soon, measuring the other two
chirally odd GPDs which enter Eq. (\ref{SR1}) will be more
challenging. Therefore, initial applications of Eq. (\ref{SR1}) will
have to rely on lattice QCD simulations.\cite{latodd}.

\section{Chirally odd GPDs and the Boer-Mulders Function}
The Boer-Mulders function $h_{1}^{\perp q}$ \cite{BM}
is similar to the Sivers function (\ref{S}) except that the
nucleon spin is replaced by the quark spin ${\bf s}$
\bea
\label{BM}
f_{q^\uparrow/p}(x,{\bf k}_\perp)&=& \frac{1}{2}\left[
f_1^q(x,{\bf k}_\perp^2)
-h_{1}^{\perp q}(x,{\bf k}_\perp^2)\frac{(\hat{\bf P}\times 
{\bf k}_\perp)
\cdot{{\bf s}}}{M}\right]
\eea
and describes the correlation between the $\perp$ momentum and
the $\perp$ spin of the ejected quark in semi-inclusive DIS
from an unpolarized target.
In Sec. \ref{sec:sivers}, a mechanism was suggested
through which the FSI in semi-inclusive DIS translates a position
space asymmetry in the target into a momentum asymmetry for the
outgoing quark. Applying the same mechanism here yields again a
negative correlation between the sign of the momentum asymmetry  
and the sign of the deformation in position space, i.e. we
expect
\be
\frac{h_{1}^{\perp q}}{2\tilde{H}_T+E_T} \sim
\frac{f_{1T}^{\perp q}}{E} < 0.
\ee
Tests of this qualitative relationship require lattice
determinations of $2\tilde{H}_T+E_T$, while $h_{1}^{\perp q}$
is accessible in polarized Drell-Yan experiments.\cite{BM}

\section*{Acknowledgments}
This work was supported by the DOE (DE-FG03-95ER40965).



\end{document}